\documentclass[12pt,a4paper]{article}

\usepackage{epsfig}
\usepackage{amstex}

\def\nostrocostrutto#1\over#2{\mathrel{\mathop{\kern 0pt \rlap 
  {\raise.2ex\hbox{$#1$}}}
  \lower.9ex\hbox{\kern-.190em $#2$}}}
\def\lsim{\nostrocostrutto < \over \sim}   %less or around ...
\def\gsim{\nostrocostrutto > \over \sim}   %greater or around...

\newcommand{\eref}[1]{(\ref{#1})}      % automatically puts ( ) around ref.s

\newcommand{\N}{{\mathcal N}}
\newcommand{\Nq}{{\mathcal N}_q}
\newcommand{\Ng}{{\mathcal N}_g}
\newcommand{\Ord}{{\mathcal O}}

\newcommand{\MeV}{{\rm MeV}}

\newcommand{\Qzpm}{Q_0}
\newcommand{\Lpm}{\Lambda}

\catcode`@=11
\newcount\@tempcntc
\def\@citex[#1]#2{\if@filesw\immediate\write\@auxout{\string\citation{#2}}\fi
  \@tempcnta\z@\@tempcntb\m@ne\def\@citea{}\@cite{\@for\@citeb:=#2\do
    {\@ifundefined
       {b@\@citeb}{\@citeo\@tempcntb\m@ne\@citea\def\@citea{,}{\bf ?}\@warning
       {Citation `\@citeb' on page \thepage \space undefined}}%
    {\setbox\z@\hbox{\global\@tempcntc0\csname b@\@citeb\endcsname\relax}%
     \ifnum\@tempcntc=\z@ \@citeo\@tempcntb\m@ne
       \@citea\def\@citea{,}\hbox{\csname b@\@citeb\endcsname}%
     \else
      \advance\@tempcntb\@ne
      \ifnum\@tempcntb=\@tempcntc
      \else\advance\@tempcntb\m@ne\@citeo
      \@tempcnta\@tempcntc\@tempcntb\@tempcntc\fi\fi}}\@citeo}{#1}}
\def\@citeo{\ifnum\@tempcnta>\@tempcntb\else\@citea\def\@citea{,}%
  \ifnum\@tempcnta=\@tempcntb\the\@tempcnta\else
   {\advance\@tempcnta\@ne\ifnum\@tempcnta=\@tempcntb \else \def\@citea{--}\fi
    \advance\@tempcnta\m@ne\the\@tempcnta\@citea\the\@tempcntb}\fi\fi}
\catcode`@=12
\begin{document}

\setcounter{page}{0}
\thispagestyle{empty}

\title{Unified QCD Description of Hadron and Jet Multiplicities}

\author{Sergio Lupia\thanks{E-mail address: lupia@mppmu.mpg.de}  
\ and \  Wolfgang Ochs\thanks{E-mail address: wwo@mppmu.mpg.de}}

\date{{\normalsize\it Max-Planck-Institut 
f\"ur Physik, (Werner-Heisenberg-Institut) \\
F\"ohringer Ring 6, 80805 M\"unchen, Germany}}

\maketitle
\thispagestyle{empty}

\begin{abstract}
The evolution equation for parton multiplicities in quark
and gluon jets which takes into account the soft gluon interference
is solved numerically using the initial conditions at threshold.
If the $k_\perp$-cutoff $Q_c$ is lowered towards the hadronic scale 
$Q_0$ of a few hundred MeV, the jets are fully resolved into hadrons. 
Both hadron and jet multiplicities in $e^+e^-$ annihilation are 
well described with a common normalization. Evidence is
presented within this perturbative 
approach that the coupling $\alpha_s(k_\perp)$
rises by an order of magnitude when approaching the low energy region.
The ratio of hadron multiplicities in gluon and quark jets is found
smaller than in previous approximate solutions of the evolution
equation.
\end{abstract}

\vspace{-15cm}

\rightline{MPI-PhT/97-46}
%\rightline{Draft 1} 
\rightline{July 18, 1997} 

\newpage

\section{Introduction}

One of the simplest characteristics of the hadronic final state is
the particle multiplicity. Following the ideas of a soft hadronization 
mechanism \cite{bcm,dkmt} the observable hadron multiplicity
in a quark or gluon jet is expected to follow closely that 
of the final QCD partons in its energy dependence. More specifically,
within the picture of local parton hadron duality (LPHD) \cite{lphd}
the primary parton of energy $E$ generates a parton cascade which
is evolved down to small scales 
of a few hundred MeV for the transverse momentum cutoff $Q_0$. Then, within 
the so-called modified leading logarithmic approximation (MLLA),  
which takes
into account the leading double logarithmic and next to leading
single logarithmic terms, 
a good description of the mean
charged particle multiplicity in $e^+e^-$ annihilation as a function
of the primary $cms$ energy $\sqrt s\equiv Q=2E$ is obtained with
a formula
\begin{equation}
\N^{e^+e^-}_{ch}(Q)=2K_{ch} \N_q(\frac{Q}{Q_0},\frac{Q_0}{\Lambda})+ {\rm const} .
\label{nch}
\end{equation}
Here $\N_q$ denotes the multiplicity of partons in a single
quark jet of energy $E$ and $\Lambda$ is the QCD scale.
From fits in the energy range
$Q=3\ldots 160$ GeV one typically finds $Q_0\gsim\Lambda\approx$
 250 MeV and the normalization $K_{ch}\approx 1.2$ \cite{lphd,opal,lo}.
Assuming $K_{ch}={2\over 3} K_{all}$ the 
number of all hadrons (charged and neutral) is then about twice
as large as the number of partons ($K_{all}\approx 2$).

Another characteristic of the final state is the mean multiplicity
of jets $\N_{jet}(Q_c,Q)$ for a given resolution parameter $Q_c$, or,  
in conventional normalization, $y_c=Q^2_c/Q^2$. 
Early results at low resolution 
in $\Ord (\alpha_s^2)$ \cite{kl} played an important role as a QCD test and 
in the determination of the running coupling  $\alpha_s$.
The calculation of jet multiplicities to all orders in $\alpha_s$
became feasible using the Durham/$k_\perp$ algorithm \cite{durham} which
takes as separation parameter for two jets the measure
\begin{equation}
y_{ij}=2\; {\rm min}\; (E_i^2,E_j^2)\; \frac{1-\cos \Theta_{ij}}{Q^2}
\label{durhamy}
\end{equation}
which for small relative angles $\Theta_{ij}$ approximates the
rescaled transverse momentum $(k_\perp^{ij})^2/Q^2$ 
of the jet of lower momentum with respect to that of higher momentum.
Then the jet
multiplicity $\N_{jet}(y_c,Q)$ refers to all jets with separation
$y_{ij}\geq y_c$. Jet multiplicities in leading and next-to%
-leading order of $\ln y_c$ have been calculated \cite{bs,cdotw,cdfw,do}
and, after matching with the full $O(\alpha_s^2)$ results, a good
description of jet multiplicities down to $y_c\sim 10^{-3}$ 
has been obtained \cite{l3jm,opaljm}. In these calculations
the jet multiplicities are derived in absolute normalization ($K_{all}=1$) from
\begin{equation}
\N^{e^+e^-}_{jet}(Q_c,Q)=2\N_q(\frac{Q}{Q_c},\frac{Q_c}{\Lambda}). 
\label{njet}
\end{equation}

In this paper we investigate whether both observables, jet
multiplicities and hadron multiplicities, can be described
in a unified way. Indeed, the resummation of the perturbative
series has been achieved in both cases by using the same type
of evolution equation. Also, the Durham/$k_\perp$ algorithm
of the jet definition (\ref{durhamy}) coincides for $Q_c\to Q_0$
with the prescription $k_\perp \geq Q_0$ 
which is applied in the description of  hadrons 
within the LPHD picture. At a first sight there is no smooth transition for 
$Q_c\to Q_0$ from (\ref{njet}) to (\ref{nch}) because of the two additional
constants in (\ref{nch}). Furthermore,  various jet
observables for small $y_c$ deviate from the perturbative
predictions (see, for example, review \cite{ko}), 
and this is often taken as evidence for nonperturbative
hadronization effects at the resolution scales $Q_c\sim 1 - 2$ GeV. 

We show 
that these problems -- in the case of the mean jet or particle multiplicity --
disappear if the underlying coupled evolution equations for quark
and gluon jets are solved with sufficiently high accuracy; here we
perform a numerical integration to obtain the exact solutions. 
%An important role is played
%by the running coupling constant which leads to a maximal
%relative separation of particle and jet multiplicities ($\N-2$) near
%threshold $y_c\to 1$.

\section{Evolution equation for multiplicities}

We start from the evolution equation for the generating functional
of the multiparton final state \cite{dok,dkmt}
which, in the parton splitting process $A\to B + C$, takes into account 
angular ordering, energy conservation 
and the running coupling at the one-loop order.  
By appropriate differentiation we obtain the evolution equations
for the mean parton multiplicities 
$\Nq$ and $\Ng$ in quark and gluon jets with
jet virtuality $\kappa$ (see also \cite{do})
or with
\begin{equation}
\eta=\ln \frac{\kappa}{Q_c}, 
\qquad \kappa=Q\sin \frac{\Theta}{2}. \label{etadef}
\end{equation}
where $\Theta$ denotes the maximum  angle between the outgoing partons $B$
and $C$. 
%For small angles $\kappa\approx E\Theta$ corresponds to the maximum
%transverse momentum in the jet.
At fixed  cutoff $Q_c$ these equations read 
\begin{eqnarray}
\frac{d\Ng (\eta)}{d\eta} &=& \int_{z_c}^{1-z_c} dz 
      \frac{\alpha_s(\tilde k_\perp)}{2\pi}[\Phi_{gg}^{asy}(z)
      \{\Ng(\eta+\ln z)+\Ng(\eta+\ln (1-z))-\Ng(\eta)\}\nonumber\\
   &\quad & \qquad +n_f \Phi_{gq}(z)
       \{\Nq(\eta+\ln z)+\Nq(\eta+\ln (1-z))-\Ng(\eta)\}] \label{eveq}
  \\ \nonumber
\frac{d\Nq (\eta)}{d\eta} &=& \int_{z_c}^{1-z_c} dz
      \frac{\alpha_s(\tilde k_\perp)}{2\pi}\Phi_{qg}(z)
      \{\Ng(\eta+\ln z)+\Nq(\eta+\ln (1-z))-\Nq(\eta)\}.
\end{eqnarray}
The splitting functions $\Phi_{AB}$ 
for parton splitting $A\to B$~\cite{dglap} are taken with normalization as in
\cite{dkmt}, i.e., $\Phi_{gg}(z) \simeq 4 N_c/z$ at small $z$ ($N_C$ and $n_f$ 
denote the number of colours and flavours). 
Because of its symmetry property it is convenient to
replace $\frac{1}{2}\Phi_{gg}(z)$ by the asymmetric function 
$\Phi_{gg}^{asy}(z)=(1-z)\Phi_{gg}(z)$ in (\ref{eveq}) \cite{do}.

The boundaries of the integral over the parton momentum fractions $z$ 
are determined by the lower cutoff 
in the transverse momentum measure $\tilde k_\perp$ 
defined according to  (\ref{durhamy}) with the approximation $z_A\approx
1-z_B$   
\begin{equation}
\tilde k_\perp  =  {\rm min}(z,1-z) \kappa\;\geq\; Q_c.  
\label{kperp}
%z_c & = & \sqrt{y_c} \; 
%\frac{Q_c}{Q} =e^{-\eta},\quad  z_c \leq \frac{1}{2} \label{zc}
\end{equation} 
The lower bound $z_c$ in (\ref{eveq})
is obtained for minimal $z=Q_c/(Q\sin  \frac{\Theta}{2})$ which is found for  
$\Theta\approx \frac{\pi}{2}$ (in this configuration, because of transverse
momentum conservation, the high energy secondary parton has  production angle
$\vartheta_A\approx 0$ and the soft particle 
$\vartheta_B\lsim  \frac{\pi}{2}$); more generally, $\Theta$ corresponds to
the half opening angle of the jet ($\vartheta_B \le \Theta$). 
Therefore $z_c$ is given by
\begin{equation}
z_c  = \frac{Q_c\sqrt{2}}{Q} =\sqrt{2 y_c} = e^{-\eta} . 
 \label{zc}
\end{equation}  
Since $z_c \leq \frac{1}{2}$, one finds 
$y_c\leq \frac{1}{8}$ and $\eta \geq \ln 2$.

%In the present application to the reaction $e^+e^- \to $ hadrons we follow
%the generally adopted convention using the full $cms$ energy $Q=2E$ as the
%scale in the definition of $y_c$ in (\ref{durhamy}) and  $k_\perp$
%in (\ref{kperp}) rather than the jet energy $E$ itself which is the natural
%scale for the single jet. As the multiplicities depend only on the ratios
%$\frac{Q}{Q_c}$ and $\frac{Q_c}{\Lambda}$ one obtains in the limit $Q_c\to
%Q_0$ values for $Q_0$ and $\Lambda$ which are twice as large as those
%obtained with scale $E$. To avoid confusion we write
%for this convention $\Qzpm ={\rm min} (Q_C)=2 Q_0$ where $Q_0$ refers to the
%scale $E=Q/2$ for the single jet as in (\ref{nch}); in the same way we
%define $\Lpm =2 \Lambda$. The solution of (\ref{eveq}) in double logarithmic
%approximation then reads
%$\N_g \sim \exp \int^{\eta_0} dy 2N_C\alpha_s(y)/\pi$
%with $\eta_0=\ln(Q/\Qzpm)=\ln(E/Q_0)$ consistent with the known result
%\cite{dkmt}. 

The coupling is given by 
$\alpha_s(\tilde k_\perp,n_f)=2\pi/(b \ln (\tilde k_\perp/\Lpm))$
with $b=(11N_C-2n_f)/3$.  
We evolve $\alpha_s(\tilde k_\perp,n_f)^{-1}$ 
with $\tilde k_\perp$ by matching
smoothly the couplings for $n_f$ and $n_f+1$ at twice 
the heavy quark thresholds according to the formula 
 \begin{equation} 
 \displaystyle\frac{1}{\alpha_s(k_t)} = 
    \displaystyle\frac{\Theta (k_t - 2 m_{n_f})}{\alpha_s^{(n_f)}(k_t)} -  
    \sum_{i=4}^{n_f} \biggl( 
   \displaystyle\frac{1}{\alpha_s^{(i)}(m_i)} - 
    \displaystyle\frac{1}{\alpha_s^{(i-1)}(m_i)} \biggr) 
    \Theta (k_t - 2 m_i) 
 \end{equation} 
 The factor two in the threshold for heavy quark production takes into account 
that heavy quarks are produced in pairs by a gluon.

The differential equations are defined only for $\eta\geq \ln 2$. 
The initial conditions for the solutions of 
(\ref{eveq}) then read (for any fixed $Q_c$)
\begin{equation}
\Ng(\eta)=\Nq(\eta)=1 \qquad {\rm for} \quad 0\leq\eta\leq \ln 2.
\label{init}
\end{equation}

Analytical solutions using the boundary conditions at threshold  
have been derived within the MLLA for
gluon jets\cite{dktint} and for the coupled system of quark and gluon
jets\cite{cdfw}. Whereas the  MLLA yields a good high energy behaviour, it leads
to inconsistencies ($\N_a'(\eta) < 0$) near threshold. In the present
analysis we solve
the equations (\ref{eveq}) and (\ref{init}) numerically.\footnote{We
 start from
$\N_a(\eta=\ln 2)=1$, calculate $\N_a(\eta+\delta \eta)$ 
in steps of length $\delta \eta$ from the
derivatives (\ref{eveq}) as $\N_a'(\eta)\delta\eta$
using the trapezoidal rule for the integration 
and linear interpolation for the required $\N$ values under the integral.
The integration is performed with logarithmic variable $y=\ln z$.}

The two jets
in $e^+e^-$ annihilation evolve independently
in this approximation. Near threshold this
factorization of the generating functional does not hold any more
and non-logarithmic terms become important. For the  $k_\perp$
resolution criterion
(\ref{durhamy}) the inelastic threshold is found for
symmetric 3-jet events ($\Theta=\frac{2\pi}{3}$, $z=\frac{2}{3}$) 
with $y_c={1\over 3}$ or $\eta=\frac{1}{2}\ln \frac{3}{2}$. 
This is to be compared with $y_c=\frac{1}{8}$ and $\eta=\ln 2$ 
for eq.~\eref{eveq}. 
The difference of thresholds will clearly affect the results in the large
$y_c$ region.

An improvement in this region
can be achieved by replacing the contribution of $\Ord (\alpha_s)$ in
(\ref{eveq}) by the explicit result for $e^+e^-\to 3$ partons in
$\Ord (\alpha_s)$ (see also Ref. \cite{cdfw,bs}). 
The lowest order contribution
$\N^{(1)}_{q}$ of the evolution equation is obtained by taking the 
first iteration of (\ref{eveq}) with the initial conditions
(\ref{init}), i.e., replacing the three $\N$-terms in the 
curly brackets in (\ref{eveq}) by unity.   
The full $\Ord (\alpha_s)$ contribution is found by numerical 
integration of
\begin{gather}
\N^{3-jet}(y_c) =2 \int_{1/2}^1dz_1\int_{1-z_1}^{z_1} dz_2 \Theta(d_{23}-y_c)
   \frac{C_F \alpha_s(\tilde k_\perp)}{2\pi} \frac{z_1^2+z_2^2}{(1-z_1)(1-z_2)}
  \label{jet3}\\
d_{23}=\min\left(\frac{z_2}{z_3},\frac{z_3}{z_2}\right)(1-z_1)>y_c
\label{d23}
\end{gather}
where $z_1$ ($z_2$) denote the quark (antiquark) and
$z_3=2-z_1-z_2$ the gluon momentum fractions ($C_F$ = 4/3). 
Here the coupling is taken again as
running with the $k_\perp$ measure (\ref{durhamy}) 
$\tilde k_\perp=(d_{23} Q^2)^{\frac{1}{2}}$ as scale in the integration
region $z_2<z_1$ in (\ref{jet3}). 
%The corresponding result for the gluon jet is taken as in  (\ref{jet3})
%but with $C_F$ replaced by $N_C$.
The $\Ord (\alpha_s)$ corrected solution $\N_{corr}^{e^+e^-}$ 
is then obtained from
\begin{equation}
\N_{corr}^{e^+e^-}(y_c) =2 \N_q(y_c)-2 \N_q^{(1)}(y_c) +\N^{3-jet}(y_c).
\label{nepem}
\end{equation}
A corresponding improved result for the gluon jet is not yet available and
has to be considered for each process separately. In this paper 
we apply the equation (\ref{nepem}) to the full range of scale parameters 
$\Qzpm<Q_c<Q$, 
i.e., we study the jet region ($y_c\gsim 0.01$) and also the
transition from the jets to the fully resolved hadrons ($y_c\sim (\Qzpm/Q)^2$).

\section{Confronting predictions with experiment}
\subsection{Multiplicities in $e^+e^-$ annihilation}

In Fig. 1a we show the data of the average jet multiplicity at $Q$ = 91 GeV 
\cite{l3jm,opaljm} as a function of the resolution parameter
$y_c= Q^2_c/Q^2$ as obtained with the $k_\perp$ algorithm. 
The theoretical predictions from (\ref{nepem}) for the jet data are given in
absolute normalization in terms of the single parameter $\Lambda$.
Also shown are
the data on hadron multiplicities in the energy range $Q$ = 1.6$\ldots$ 
91 GeV~\cite{hadron}
taken as $\N_{all}=\frac{3}{2} \N_{ch}$ 
as a function of the
same scale parameter, now calculated as  $y_c=Q^2_0/Q^2$,  where 
the parameter  $Q_0$ corresponds to the parton $k_\perp$
cutoff characterizing a hadronic scale  and is
obtained from a fit to the data. Another adjustable parameter here is the
overall normalization $K_{all}$ which relates the parton and hadron
multiplicities in $\N_{all}=K_{all} \N_{corr}^{e^+e^-}$. 
We determine first the $\Lambda$ parameter from the jet multiplicity
(lower data set in Fig. 1) and then the $K_{all}$ and $Q_0$ from the
hadronic multiplicity (upper data set). A good description of the data 
is obtained with parameters 
\begin{equation}
K_{all}=1, \qquad \Lambda=500\pm 50~ \MeV 
\qquad  \lambda=\ln \frac{Q_0}{\Lambda}=0.015\pm 0.005  \label{results}
\end{equation}
which correspond to the curves shown in Fig. 1a. The quantity $\N-2$ derived  
from the same experimental data is shown in Fig. 1b. The dashed curves
represent the contribution $2\N_q$ in (\ref{nepem}) from the evolution
equation alone. The fits can be seen to describe  quantitatively the
hadron multiplicity, the jet multiplicity for the small and large $y_c$,
whereas in the intermediate region around $y_c \simeq 0.01$ 
the fit exceeds the data on $\N-2$ by up to  20\% (up to 5\% in $\N$). 
Deviations of this type may indicate
the importance of two-loop $\Ord (\alpha_s^2)$ terms not included in
the analysis.  

As a remarkable result of our analysis we find that the common normalization
 $K_{all}=1$ is possible without difficulty; 
the normalization parameter $K_{all}$ is correlated with $Q_0$
and can be varied within about 30\%.
The solution with $K_{all}$ = 1 is natural as 
it provides  the correct boundary conditions $\N=2$
at threshold for both hadrons and jets. Then
a continuous connection  between jet and hadron
data results, since both are described by  
the same evolution equation (\ref{nepem}). In this description 
a single hadron corresponds to a single parton of low virtuality
$Q_0$.

It is interesting to compare the two sets of data at the same
$y_c$. The difference between the two curves comes entirely from
the running of $\alpha_s$. Namely, if $\alpha_s$ were kept
fixed in equations (\ref{eveq}) the resulting multiplicities
would depend only on the ratio of the available scales through the variable
$\eta = -\frac{1}{2}\ln (2 y_c)$ 
but not on the absolute size of $Q$ and the two curves in Fig. 1 would
coincide. 
A model with fixed $\alpha_s$ would predict at high energy a power like
dependence on Q, i.e. a straight line in Fig. 1a between the two curves
shown.

The largest difference occurs near the inelastic threshold
at $y_c\approx {1\over 8}$ or $\eta\approx \ln 2$. In this region 
$\N-2$ is dominated by the contribution of order $\alpha_s$ 
and the jet and hadron results at the same $y_c$ are in the ratio of the 
relevant typical coupling constants. At the lowest available energy 
for the hadron multiplicity ($y_c \sim 0.1$, $Q=1.6$ GeV) 
this ratio of couplings is larger than 10, 
 so the typical coupling to produce hadrons at 1.6 GeV is 
$\alpha_s \gsim 1$. The good matching of the prediction for hadrons with
the boundary condition at threshold suggests that the coupling is rising
even more towards lower energies. 
The strong variation of the coupling at small scales has been found 
important also in the
description of particle energy spectra\cite{lo,klo}. 
On the other hand, the jet multiplicity  is first
rising very slowly with decreasing $y_c$, because of the small coupling 
($\alpha_s\sim 0.1$). Only 
for very high resolution, if $Q_c$ is lowered from 900 MeV 
($y_c\sim 10^{-4}$) to the final 500
MeV, about three quarters of the final multiplicity are 
 produced, three times
as many jets as in the large complementary kinematic region.
 The steep rise toward small $y_c$
 in Fig.~1 reflects the singularity in the coupling 
 at $y_\Lambda=\Lambda^2/Q^2$, which is 
 however screened by the hadronization scale  $y_0=(\Qzpm)^2/Q^2\gsim y_\Lambda$.   
This behaviour is qualitatively described by the
high energy DLA result\cite{dkmt} $\N\sim
K_0(A\sqrt{\lambda_c})I_1(A\sqrt{\ln(\kappa/\Lambda)})+\ldots$ with
$A=\sqrt{16N_C/b}$ which at high energies and small $\lambda_c = \ln
(Q_c/\Lambda)$ behaves as $\ln \lambda_c \exp (A \sqrt{\ln (\kappa/\Lambda)})$.
This form, for fixed $Q_c$, describes the slow rise of multiplicity with  
$\kappa \sim Q$, whereas, for fixed $Q$, it 
exhibits the logarithmic singularity for $\lambda_c \to$ 0. 
This singularity arises in the first iteration of (\ref{eveq}) 
(the ``Born-term'') from the contribution 
$\int_0^\eta dy/(y+\lambda)$.

Previous approximate solutions had been restricted to $y_c \gsim 0.002$ for
jets, whereas for hadrons the analytical results could  only be 
applied to the higher energy region with a larger $K_{ch}$ 
factor.\footnote{The different value  of the
normalization factor as compared
to previous analyses (for example \cite{lphd,opal,lo}) 
is due to the use in the MLLA approximation of the initial conditions
which have negative slope $\N'(\eta=0)<0$ and therefore lead to a delayed
rise with energy~\cite{cdfw}. Our exact solution in the present 
$cms$ energy range is about twice as large and rises more slowly than 
the analytical MLLA solution for $Q_0 = \Lambda$~\cite{dktint}.}
Our results are consistent with the previous finding $\lambda \lsim 0.1$ 
 from the energy moments \cite{lo}; 
on the other hand, $\lambda$ = 0 is not allowed in~\eref{eveq},  
as the multiplicity would diverge in this case.

It will be interesting to study this behaviour further near the
transition $Q_c \to \Qzpm$ also at other $cms$ energies, especially for the jet 
multiplicity, since it varies strongly in the low $y_c$ range. 
In the experimental analysis using eq.~\eref{durhamy} the full hadronic
multiplicity will only be reached for $y_c \to$ 0. The same lower cutoff for
partons and hadrons is achieved if we interpret $Q_c$ as transverse mass
cutoff with $Q_0$ as effective mass parameter (see also \cite{lo,klo}), 
i.e., if we relate the theoretical predictions and experimental results from 
cutoff~\eref{durhamy} by $y_c^{th} = y_c^{exp} + Q_0^2/Q^2$. 
%If we follow the rise of jet multiplicity with decreasing scale $y_c$
%it is interesting to observe that three quarter of  the jets (particles) are
%produced in the final narrow regime of strong coupling below $y_c=10^{-4}$  
%($\Qzpm\approx 0.5~ \GeV\lsim Q_c
%\lsim 0.9$ GeV)

The value~\eref{results} for the scale $\Lambda$ is larger than that found in
previous analyses of particle spectra\cite{lphd,opal,lo}. This difference
comes from taking the Durham $\tilde k_\perp$ not only in the integral
boundaries but also as argument of $\alpha_s$ and in
the definition of hadrons by $\tilde k_\perp \ge Q_0$. We performed an
alternative calculation where we used the standard $k_\perp < \tilde k_\perp$ 
as argument of
$\alpha_s$ (i.e., $k_\perp = z (1-z) \kappa$ instead of eq.~\eref{kperp} and the 
exact $k_\perp$ in eq.~\eref{jet3}) and inserted the cutoff $\Theta(k_\perp - Q_0)$ into
eqs.~\eref{eveq} and ~\eref{jet3}; then a description of comparable quality is
obtained with the lower scale $\Lambda \sim 0.35$ GeV and the same $\lambda$ 
parameter\footnote{The remaining difference to the earlier $\Lambda \simeq$ 250
GeV comes from taking the scale $\kappa = \sqrt{2} E$ in~\eref{kperp} instead of
$\kappa = E$.}.  
The threshold for hadrons is shifted in this case to the lower value $y_c =
1/12$ and also the hadron data points are shifted to the left in Fig.~1
according to the lower value of $Q_0$. 
Our conclusions about the order of magnitude variation of $\alpha_s$ remain
unaltered. 
The ambiguity in choosing the scale for $\alpha_s$ may be reduced in 
2-loop calculations; for now we stay
with the conceptually simplest possibility and take $\tilde k_\perp$
throughout. 

\subsection{Multiplicities in gluon jets}
The evolution equations (\ref{eveq}) yield also results on the mean
multiplicity $\N_g$ in gluon jets. The results derived here 
refer to the multiplicity in the
full hemisphere of the gluon jet and can be measured, for example, in 
a final state with a primary $gg$ colour singlet state. Such a state is
expected to be produced in the decay of heavy quarkonia; recently the
multiplicity in the decay $Y\to gg\gamma$ has been measured by the CLEO
Collaboration \cite{cleo}. An initial state of this type can also be
realized approximately in $e^+e^-\to 3$ jets with nearly antiparallel $q$
and $\overline q$ recoiling against the gluon \cite{dkmt,gary};  this
configuration has been analysed by the OPAL Collaboration
\cite{opalglu}\footnote{The full multiplicity in the hemisphere opposite to
$q\bar q$ corresponding to our calculation 
is obtained if the higher jet multiplicities are also taken into
account.}.
Our results do not apply immediately to other configurations (like ``Y'' or
``Mercedes'') where the large angle soft radiation from the gluon is suppressed
 (for reviews, see \cite{fuster,ko}).

Of special interest is the ratio of gluon and quark jet multiplicity 
$r=\N_g/\N_q$, which is predicted to approach 
asymptotically the value $r=\frac{9}{4}$ \cite{bg}.
The next-to-leading order corrections decrease this ratio 
at finite energies \cite{mw} and yet further with the inclusion of higher order
corrections \cite{gm,dremin}. There is an uncertainty with these 
asymptotic predictions in that the region of validity is not known a priori.
A complete solution of the evolution equations (\ref{eveq}) requires
an initial value for the multiplicities at some energy. 
Such results with absolute
normalization at threshold have been presented within the MLLA \cite{cdfw}.

We have obtained 
the results from our evolution equation (\ref{eveq}) with initial condition
(\ref{init}), however, we have not yet included a low energy correction term
as in (\ref{nepem}) which is beyond the scope of this paper. The results
for the ratio $r$
are shown in Fig. 2 again as function of $y_c=(Q_c/Q)^2$ for hadron
multiplicities in comparison with the OPAL and 
CLEO data mentioned above. Also shown is the prediction for jets at fixed
$Q=91$ GeV which could be tested using the same type of events with
antiparallel $q\bar q$. 

There is a very good agreement of our calculation with the OPAL data
which refer to jet energies of $\sim$40 GeV. From Fig. 1b one
may expect that the inclusion of the full $\Ord (\alpha_s)$ corrections 
would not modify the prediction essentially (by more than about 10\%).
On the other hand, the CLEO data fall considerably below the
prediction $r\sim 1.25$ from the evolution equation.
In this region, however, the low energy corrections may not be small:
in Fig 1b these corrections correspond to a factor of two. A more precise
prediction of the $Y$ data will require the inclusion of the 
full $\Ord (\alpha_s)$ correction to the $Y \to gg\gamma$ process. 
%In the low energy region where the configurations with large relative angles 
%dominate  the additional gluon could be produced from the heavy 
%quark and not from the gluon, as naively expected; then the rate would be
%proportional to $C_F$ and not to $N_C$ resulting in a smaller value of $r$. 

\section{Conclusions}
We solved the QCD evolution equations for quark and gluon jets 
exactly by numerical integration and also supplemented the full $\Ord
(\alpha_s)$ correction for $e^+e^-$ annihilation. 
With this improved accuracy beyond MLLA we obtain a unified
description with  common absolute normalization
of the mean jet multiplicity  at LEP-1 and the hadron
multiplicities in the energy range from $1.6$ to 91 GeV 
within the errors or at least within 5\%. In this description only the QCD
scale $\Lambda$ and a $k_\perp$ cutoff parameter $Q_0\gsim\Lambda$ 
for the hadrons according to the LPHD picture have been adjusted to the
data. The ratio of gluon to quark multiplicity at LEP-1 is described as well.

Of particular interest is the success of this model in the regions where
perturbative QCD is not expected to be relevant a priori, namely for small
transverse momenta, where the running coupling gets large, 
either for the hadron multiplicity near threshold or for the jet multiplicity 
at very high resolution. In both cases the strong 
effect of the running $\alpha_s$ is clearly seen from the comparison
of the two multiplicities 
at the same scale $y_c$; taking the transverse momentum  as the
argument of $\alpha_s$ provides about the right separation. 
It appears that perturbative QCD is still
successful in these extreme kinematic regions for this inclusive quantity.
 
The model for hadron production emerging from this analysis is very simple.
A hadron can be treated formally 
like a narrow microjet of partons but with the radiated 
partons being so soft ($\Lambda<k_\perp<Q_0$) that they cannot be resolved
and are therefore confined in a region characterized by the 
hadronic scale $Q_0$ of a few hundred MeV. So a hadron in this picture 
does not correspond to a colour singlet state of \lq\lq valence partons"
but rather to a single parton (quark or gluon) with about the same momentum 
together with a \lq\lq sea" of soft confined partons which take care of the
colour neutralization. In this way the soft hadronization picture
can become consistent with the perturbative treatment of particle production.

\newpage

\section*{Figure Captions}

{\bf Fig. 1a}: 
Data on the average jet multiplicity at $Q$ = 91 GeV\cite{l3jm,opaljm} and 
the average hadron multiplicity (assuming $\N = \frac{3}{2} \N_{ch}$) 
at different $cms$
energies\cite{hadron} with $Q_c$ = 0.507 GeV as a function of $y_c$. 
The solid (dashed) line shows the prediction in absolute normalization 
 for the hadron (jet) 
multiplicity obtained by using eq.~\eref{nepem} with parameters from 
%$K_{all}$ = 1, $\Lambda$ = 0.5 GeV and $\lambda = 0.015$ (see also 
eq.~\eref{results}. 
The right most data point for hadrons ($Q_0$ = 1.6 GeV) refers to pions only.

\medskip\noindent 
{\bf Fig. 1b}: 
Data as in Fig. {\bf 1a}, but ${\mathcal N} - 2$ is now shown. 
The solid lines show the complete predictions  of eq.~\eref{nepem} 
for both the particle and jet multiplicities,
 the dashed lines the contribution of eq.~\eref{eveq} alone. 
The same parameters as in Fig. {\bf 1a} are used.

\medskip\noindent 
{\bf Fig. 2}: 
Data on the ratio of average hadron multiplicities in quark and gluon
jets at $y_c \sim 3 \times 10^{-5}$ from OPAL\cite{opalglu} and around $y_c \sim 
0.01$ from CLEO\cite{cleo} 
 with $Q_c$ = 0.507 GeV as a function of $y_c$. 
The solid (dashed) line shows the prediction for the ratio of 
hadron (jet) multiplicities in quark and gluon jets 
 obtained by using eq.~\eref{eveq} 
with the same parameters as in Fig.~{\bf 1}.

 \newpage

\begin{figure}[p]
\begin{center} 
\mbox{\epsfig{file=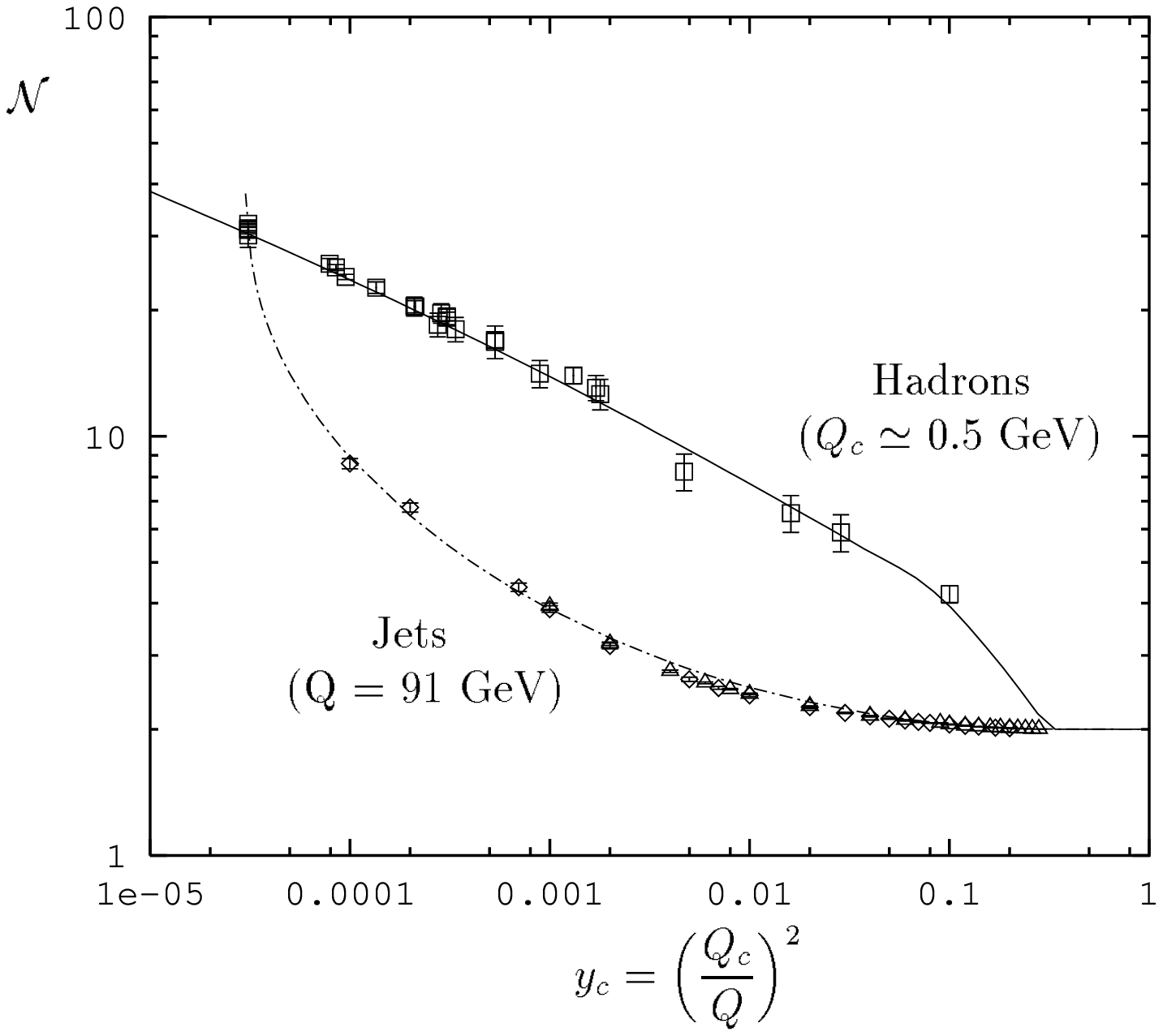,width=17cm,bbllx=4.cm,bblly=6.cm,bburx=21.cm,bbury=22.cm}}
\end{center}
\caption{a} 
\label{fig1a}
\end{figure}

 \newpage

\setcounter{figure}{0}

\begin{figure}[p]
\begin{center} 
\mbox{\epsfig{file=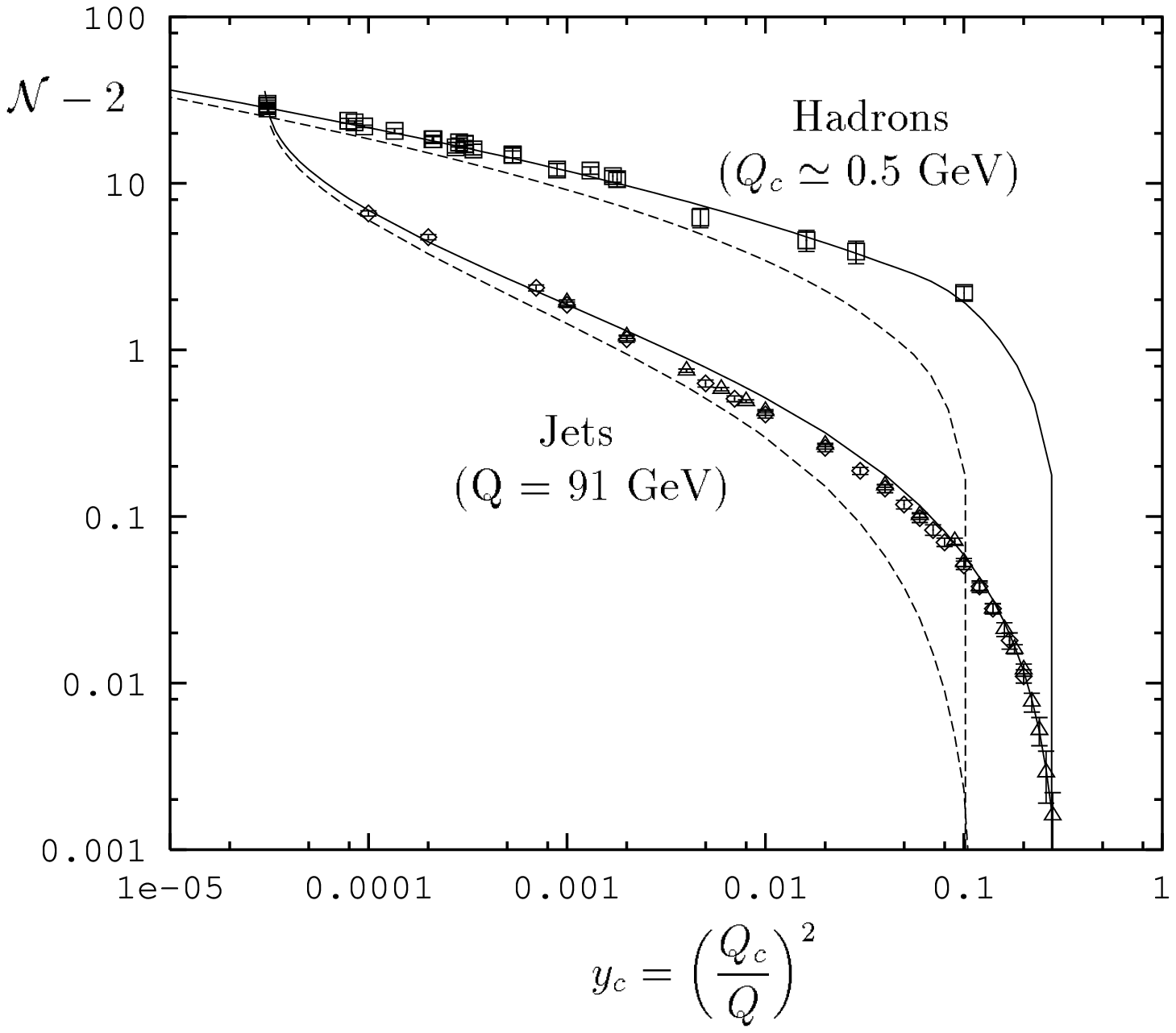,width=17cm,bbllx=4.cm,bblly=6.cm,bburx=21.cm,bbury=22.cm}}
\end{center}
\caption{b} 
\label{fig1b}
\end{figure}

 \newpage

\begin{figure}[p]
\begin{center} 
\mbox{\epsfig{file=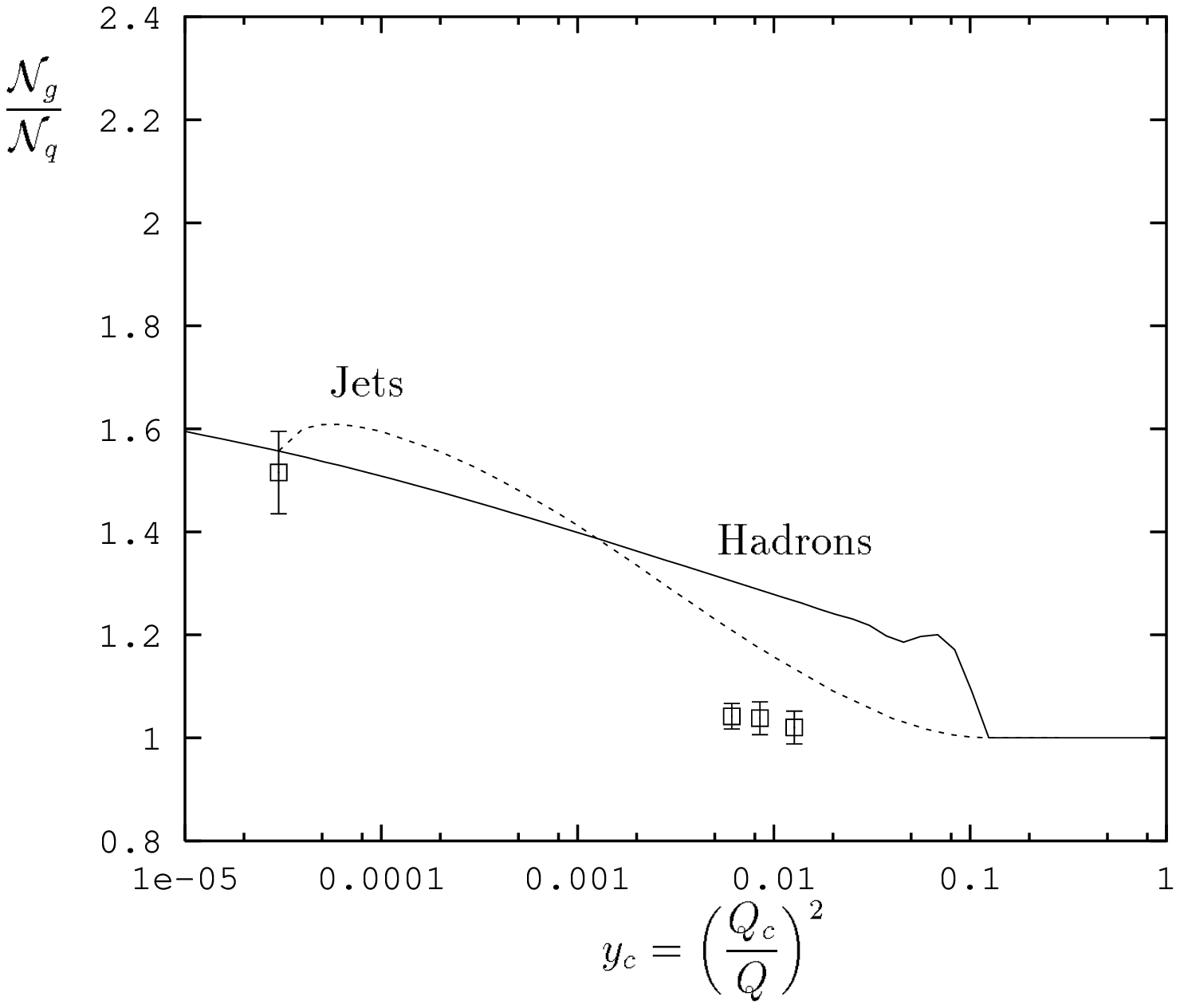,width=17cm,bbllx=4.cm,bblly=6.cm,bburx=21.cm,bbury=22.cm}}
\end{center}
\caption{} 
\label{figratio}
\end{figure}

\end{document}